\documentclass[doublecol]{epl2}

\usepackage{amsmath}
\usepackage[mathscr]{eucal}
\usepackage{dsfont}
\providecommand{\ket}[1]{\left\lvert #1\right\rangle}
\providecommand{\bra}[1]{\left\langle #1\right\rvert}
\providecommand{\bracket}[3]{\left<#1\vphantom{#2#3}\right|#2\left|#3\vphantom{#1#2}\right>}%
\providecommand{\ketbra}[2]{\ket{#1\vphantom{#2}}\bra{#2\vphantom{#1}}}

\providecommand{\omegaab}{g_{\mathrm{A}\mathrm{B}}}%
\providecommand{\omegaba}{g_{\mathrm{B}\mathrm{A}}}%
\providecommand{\abtimesba}{g_{\mathrm{A}\mathrm{B}}g_{\mathrm{B}\mathrm{A}}}%
\providecommand{\abfracba}{\frac{g_{\mathrm{A}\mathrm{B}}}{g_{\mathrm{B}\mathrm{A}}}}%
\providecommand{\bafracab}{\frac{g_{\mathrm{B}\mathrm{A}}}{g_{\mathrm{A}\mathrm{B}}}}%
\providecommand{\raizabba}{\sqrt{g_{\mathrm{A}\mathrm{B}}g_{\mathrm{B}\mathrm{A}}}}%
\providecommand{\qA}{q_\mathrm{A}}%
\providecommand{\pA}{p_\mathrm{A}}%
\providecommand{\qB}{q_\mathrm{B}}%
\providecommand{\pB}{p_\mathrm{B}}%

\providecommand{\e}{\mathrm{e}}% exponential
\providecommand{\I}{\mathrm{i}}% imaginary unit
\title{Non-hermitian model for resonant cavities coupled by a chiral mirror}
\author{R. B. B. Santos\thanks{E-mail: \email{rsantos@fei.edu.br}}}
\shortauthor{R. B. B. Santos}

\institute{
  Departamento de F\'{\i}sica, Centro Universit\'{a}rio da FEI\\
  Avenida Humberto de A. C. Branco 3972\\
  09850-901 S\~{a}o Bernardo do Campo, SP, Brazil \\
}

\abstract{Inspired by a recently observed asymmetry in the transmission of circularly polarized light through a metamaterial, we present a non-hermitian $\mathscr{PT}$-symmetric quantum model to describe the interaction of the light fields in two resonant cavities coupled via a $2D$-chiral mirror. We compute the time evolution of the light fields in this model, find two sets of operators compatible with the hamiltonian in a delocalized representation, discover the energies of the system and show that the transmission probability predicted by the model is indeed asymmetric.}

\begin{document}

\maketitle

\section{Introduction}

Recently, a remarkable $4\,\mathrm{dB}$ asymmetry was observed in the transmission of circularly polarized electromagnetic waves through an array of asymmetric split rings~\cite{Plum+Fedotov+Zheludev:2011}. This implies that the total transmission probability for circularly polarized waves incident on the front face of the array is very different from the total transmission probability for circularly polarized waves incident on the back face of the array. What would happen if this metamaterial array acted as a chiral semi-transparent mirror coupling two resonant cavities? We would expect the time evolution of the field in each cavity to be different from what is observed in the case of two cavities coupled through a regular mirror, but how different?

In this paper, we propose a non-hermitian model that describes the quantum behaviour of the light field present in two resonant cavities coupled by such a $2D$-chiral mirror. Firstly, we will resume some essential facts about optical cavities coupled by a standard, reciprocal interaction. Then, we will present our model for a non-reciprocal interaction between coupled cavities and discuss some of its consequences.

\section{Coupled cavities with reciprocal interaction}
\label{sec:Coupled_cavities}

The description of one particular mode of the electromagnetic field in a cavity is equivalent to the description of a quantum oscillator. In this paper, we consider a very simple system consisting of two identical coupled quantum oscillators ($A$ and $B$) with free ground-state energy $\hbar\omega_0/2$, described in the rotating-wave approximation by the hamiltonian
\begin{equation}\label{eq:H1}
 H = H_0 + H_\mathrm{I,h}
\end{equation}
with a free hamiltonian
\begin{equation}\label{eq:H_01}
 H_0 = \hbar\omega_0\left(a^\dagger a + b^\dagger b + 1\right)
\end{equation}
and an interaction hamiltonian
\begin{equation}
 H_\mathrm{I,h} = -\hbar g\left(a b^\dagger +a^\dagger b\right).
\end{equation}
Each quantum oscillator represents the light field in one of the cavities. The coupling between the cavities is represented by the interaction hamiltonian that describes the reciprocal interaction between two ordinarily coupled cavities. For definiteness, we choose $g>0$.

In terms of the quadrature operators $(\qA, \pA)$ for cavity $A$, and $(\qB, \pB)$ for cavity $B$, the Dirac operators are given by
\begin{align}
 a &= a(t) = \frac{1}{\sqrt{2\hbar}}(\qA + \I\pA)
 &
 b &= b(t) = \frac{1}{\sqrt{2\hbar}}(\qB + \I\pB)\\
 a^\dagger &= a^\dagger(t) = \frac{1}{\sqrt{2\hbar}}(\qA - \I\pA) %
 &
 b^\dagger &= b^\dagger(t) = \frac{1}{\sqrt{2\hbar}}(\qB - \I\pB). %
\end{align}
We assume that the lowering operators $a$ and $b$, and the raising operators $a^\dagger$ and $b^\dagger$ all obey the usual equal-time canonical commutation relations
\begin{equation}\label{eq:ab-commutation1}
 [a, a^\dagger] = 1 = [b, b^\dagger].
\end{equation}
We also assume that, at $t=0$,
\begin{align}
 a_0 \ket{n_A, n_B} &= \sqrt{n_A}\ket{n_A - 1, n_B} \\
 b_0 \ket{n_A, n_B} &= \sqrt{n_B}\ket{n_A, n_B - 1} \\
 a^\dagger_0 \ket{n_A, n_B} &= \sqrt{n_A + 1}\ket{n_A + 1, n_B} \\
 b^\dagger_0 \ket{n_A, n_B} &= \sqrt{n_B + 1}\ket{n_A, n_B +1}
\end{align}
where the ket $\ket{n_A, n_B}$ describes the state of the system in terms of the number of photons present in each cavity.

The evolution of the light field is ruled by Heisenberg equations for the Dirac operators:
\begin{align}\label{eq:coupled-i1}
 \frac{\upd a}{\upd t} &= \frac{1}{\I\hbar}[a,H] = -\I\left(\omega_0 a - g b\right) \\
 \frac{\upd b}{\upd t} &= \frac{1}{\I\hbar}[b,H] = -\I\left(-g a + \omega_0 b\right) \\
 \frac{\upd a^\dagger}{\upd t} &= \frac{1}{\I\hbar}[a^\dagger,H] = \I\left(\omega_0 a^\dagger - g b^\dagger\right) \\
 \label{eq:coupled-f1}
 \frac{\upd b^\dagger}{\upd t} &= \frac{1}{\I\hbar}[b^\dagger,H] = \I\left(-g a^\dagger + \omega_0 b^\dagger\right).
\end{align}
which may be solved by introducing lowering ($\alpha^-$ and $\beta^-$) and raising ($\alpha^+$ and $\beta^+$) operators for the two intercavity field modes by means of the canonical transformation
\begin{align}
\label{eq:transformation-i1}
 \alpha^- &= \frac{ a +  b}{\sqrt{2}} &
 \beta^- &= \frac{ a -  b}{\sqrt{2}} \\
 \alpha^+ &= \frac{ a^\dagger +  b^\dagger}{\sqrt{2}} &
 \beta^+ &= \frac{ a^\dagger -  b^\dagger}{\sqrt{2}}.
\label{eq:transformation-f1}
\end{align}
It should be remarked that the intercavity field modes are delocalized, consisting in a superposition of localized cavity modes from both cavities. It is a simple task to show that
\begin{align}
\label{eq:alphabeta-i1}
 \alpha^- &= \alpha^-_0 \e^{-\I\omega_\alpha t} & \beta^- &= \beta^-_0 \e^{-\I\omega_\beta t} \\
 \alpha^+ &= \alpha^+_0 \e^{\I\omega_\alpha t} &
 \beta^+ &= \beta^+_0 \e^{\I\omega_\beta t}.
\label{eq:alphabeta-f1}
\end{align}
where
\begin{equation}
 \omega_\alpha = \omega_0 - g \qquad \omega_\beta = \omega_0 + g
\end{equation}
are the eigenfrequencies of the intercavity field modes $\alpha$ and $\beta$, respectively. Hence, the Dirac operators are given by
\begin{align}
\label{eq:ab-i1}
 a &= \left(\cos(g t) a_0 +\I\sin(g t) b_0 \right)\e^{-\I\omega_0 t}  \\
 b &= \left(\I\sin(g t) a_0 + \cos(g t) b_0 \right)\e^{-\I\omega_0 t} \\
 a^\dagger &= (\cos(g t) a^\dagger_0 -\I\sin(g t) b^\dagger_0 )\e^{\I\omega_0 t}  \\
 b^\dagger &= (-\I\sin(g t) a^\dagger_0 + \cos(g t) b^\dagger_0 )\e^{\I\omega_0 t},
\label{eq:ab-f1}
\end{align}
while the photon number operator for cavities $A$ and $B$ are
\begin{multline}\label{eq:Na1}
 N^\mathrm{A} = \cos^2(g t)a^\dagger_0 a_0 + \sin^2(g t)b^\dagger_0 b_0 \\
  +\I\cos(g t)\sin(g t)(a^\dagger_0 b_0 - a_0 b^\dagger_0),
\end{multline}
and
\begin{multline}\label{eq:Nb1}
 N^\mathrm{B} = \sin^2(g t)a^\dagger_0 a_0 + \cos^2(g t)b^\dagger_0 b_0 \\
  +\I\cos(g t)\sin(g t)(- a^\dagger_0 b_0  + a_0 b^\dagger_0),
\end{multline}
respectively.

Time evolution of the system is controlled by the unitary operator $\mathscr{U}=\exp(-\I Ht/\hbar)$. In the delocalized basis,
\begin{equation}
 \mathscr{U}= \exp(-\I E_{n_\alpha, n_\beta} t) \ketbra{n_\alpha, n_\beta}{n_\alpha, n_\beta}
\end{equation}
where $E_{n_\alpha, n_\beta} = \hbar\left(\omega_\alpha n_\alpha + \omega_\beta n_\beta + \omega_0 \right)$. In the localized basis, however, the time evolution operator is not diagonal, leading to transitions between localized states. Since $[H_0, H_\mathrm{I,h}] = 0$, the time evolution operator may be written as an operator product,
\begin{multline}
 \mathscr{U}=\exp(-\I H_0 t/\hbar)\exp(-\I H_\mathrm{I,h} t/\hbar)\\ \approx\exp(-\I H_0 t/\hbar)\left(1 - \I  H_\mathrm{I,h} t/\hbar\right),
\end{multline}
where the first order approximation suffices for the one-photon exchange between cavities. Apart from a phase factor contributed by $\exp(-\I H_0 t/\hbar)$,
\begin{multline}
 \bracket{n'_a, n'_b}{\mathscr{U}}{n_a, n_b}\approx\delta_{n'_a, n'_b}^{n_a, n_b} \\ + \I t g \Bigl(\sqrt{n_a(n_b+1)}\delta_{n'_a, n_a-1}^{n'_b, n_b+1} + \sqrt{(n_a+1)n_b}\delta_{n'_a, n_a+1}^{n'_b, n_b-1}\Bigr),
\end{multline}
characterizing a reciprocal interaction for the one-photon exchange probability between cavities $A$ and $B$. In the following sections, we present a model in which the two cavities are coupled by a non-reciprocal mirror and highlight some of its consequences.

\section{Description of the model}
\label{sec:Description_of_the_model}

The model we are proposing to describe the interaction of two cavities coupled by a non-reciprocal interaction is a variant of the coupled cavities model presented in section~\ref{sec:Coupled_cavities} in which we change the interaction hamiltonian to
\begin{equation}\label{eq:H_I}
 H_\mathrm{I} = -\hbar\left(\omegaab a b^\dagger + \omegaba a^\dagger b\right).
\end{equation}
In this model, each quantum oscillator represents the light field in one of the cavities. The coupling between the cavities is represented by the interaction hamiltonian, which seeks to describe the transmission asymmetry aspects of the action of a $2D$-chiral mirror. This interaction hamiltonian is the simplest extension of the hermitian operator $H_\mathrm{I,h}$
that describes the reciprocal interaction between two ordinarily coupled cavities. We will restrict the present treatment to a closed system approach, disregarding any interaction of the cavities with a reservoir. The interaction of the cavities with matter, and with a reservoir will be dealt with elsewhere.

The hamiltonian $H=H_0+H_\mathrm{I}$ is a hermitian operator only if the condition $\omegaab=\omegaba^\ast$ is satisfied. In any other case, this hamiltonian will be non-hermitian. Particularly interesting is the case in which $\omegaab , \omegaba\in\mathds{R}$. In this case, even if $\omegaab\neq\omegaba$, the hamiltonian $H$ is a  $\mathscr{PT}$-symmetric operator. In this non-reciprocal case, transmission probabilities are expected to be asymmetrical since $\omegaab\neq\omegaba$.

Parity (space inversion) $\mathscr{P}$ and time reversal $\mathscr{T}$ are the basic operations of $\mathscr{PT}$ symmetry. Under a parity transformation, $q\rightarrow -q$, and $p\rightarrow -p$ while $q\rightarrow q$, $p\rightarrow -p$, and $\I\rightarrow -\I$ under the anti-linear time reversal transformation. Besides commuting with each other, the operators $\mathscr{P}$ and $\mathscr{T}$ that realize parity and time reversal transformations also satisfies $\mathscr{P}^2 = \mathds{1} = \mathscr{T}^2$. Since $\mathscr{P}$ is unitary and $\mathscr{T}$ is anti-linear, the combined $\mathscr{PT}$ operator is an antiunitary operator~\cite{Wigner:1959}. Under a $\mathscr{PT}$ transformation, $q\rightarrow -q$, $p\rightarrow p$, and $\I\rightarrow -\I$. Hence, under a spacetime inversion operation, Dirac operators just reverse their signs, that is $a,a^\dagger\rightarrow -a,-a^\dagger$.

$\mathscr{PT}$-symmetric models have been receiving a great deal of attention since Bender and Boettcher~\cite{Bender+Boettcher:1998} discussed a class of non-hermitian $\mathscr{PT}$-symmetric hamiltonian with real and positive eigenvalues. However, a $\mathscr{PT}$-symmetric hamiltonian alone is not enough to guarantee a real spectrum. In fact, if the eigenvectors of the hamiltonian are not simultaneously eigenvectors of the $\mathscr{PT}$ operator, then the lowest pair of eigenvalues dissolves into complex conjugates, the spectrum ceases to be real and the $\mathscr{PT}$ symmetry of the hamiltonian is said to be broken~\cite{Bender+Boettcher+Meisinger:1999}. This spontaneous $\mathscr{PT}$ symmetry breaking, and the \emph{exceptional points}~\cite{Kato:1995, Heiss:2004} or the \emph{non-hermitian degeneracies}~\cite{Berry:2004} at which the symmetry breaking occurs have been the subject of intense research in recent years. Even optical valves and cloaking devices based on $\mathscr{PT}$ symmetry breaking were recently proposed~\cite{Ramezani+etal:PRA2010, Lin+etal:2011} and analysed~\cite{Longhi:JPhysA2011}. A few years after the pioneer work by Bender and Boettcher, Mostafazadeh~\cite{Mostafazadeh:2003b} showed that exact $\mathscr{PT}$-symmetry is equivalent to hermiticity with respect to a suitably defined inner product. Recent reviews~\cite{Bender:2007, Mostafazadeh:2009} provide a detailed exposition of the status of the theory. Experimental realizations of systems with $\mathscr{PT}$-symmetry appeared much more recently, initially in the optics domain~\cite{Klaiman+Moiseyev+Gunther:2008, Guo+etal:2009, Ruter+etal:2010} following a theoretical analysis of coupled optical $\mathscr{PT}$-symmetric structures involving a balanced gain-loss profile~\cite{El-Ganainy+etal:OptLett2007}. More recently, an active RLC circuit realization was demonstrated~\cite{Schindler+etal:2011}. Despite the impressive results achieved in these experiments, they all share the common trait of appearing to be strictly classical systems. However, a theoretical analysis~\cite{Schomerus:2010, Yoo+Sim+Schomerus:2011} based on the scattering formulation~\cite{Gruner+Welsch:1996, Beenakker:1998, Schomerus+etal:2000} of the input-output formalism~\cite{Collett+Gardiner:1984, Gardiner+Collett:1985} in quantum optics showed that quantum noise, losses through leakage, and instability are critical to comprehend the observed behaviour of these systems in which non-hermiticity arises from coupling lossy and active elements, that is, absorbing and amplifying regions, via a reciprocal coupling. In our model, however, the source of non-hermiticity is the presence of a non-reciprocal coupling between two otherwise ordinary resonant cavities. Non-reciprocal Bragg scattering from a photonic lattice with $\mathscr{PT}$ symmetry was predicted~\cite{Longhi:PRL2009}, owing to a breakdown of invariance of the diffracted beam intensities under crystal inversion~\cite{Keller+etal:1997, Berry:1998}. Still relating to $\mathscr{PT}$ symmetry in optical cavities, Longhi~\cite{Longhi:PRA2010} built on the ideia of a coherent perfect absorber~\cite{Chong+etal:2010} and used a classical scattering formalism to show that a $\mathscr{PT}$-symmetric optical medium in a resonant cavity may behave both as a laser oscillator, and as a coherent perfect absorber. Another interesting development is the search for optical realisations of relativistic non-hermitian systems~\cite{Longhi:PRL2010}. We believe that the system analysed in this paper can be realised combining standard techniques from cavity quantum electrodynamics and metamaterials research. Therefore, it could be one of the simplest experimental realizations of a $\mathscr{PT}$-symmetric quantum system.

In the following sections, we will explore the consequences of this non-hermitian hamiltonian, owing to the non-reciprocal nature of the interaction in the case  $|\omegaab| \neq |\omegaba|$. This case is clearly impossible in a hermitian model.

\section{Equations of motion}
\label{sec:equations_of-motion}

The dynamics of the system is ruled by Heisenberg equations for the evolution of the Dirac operators:
\begin{align}\label{eq:coupled-i}
 \frac{\upd a}{\upd t} &= \frac{1}{\I\hbar}[a,H] = -\I\left(\omega_0 a - \omegaba b\right) \\
 \frac{\upd b}{\upd t} &= \frac{1}{\I\hbar}[b,H] = -\I\left(-\omegaab a + \omega_0 b\right) \\
 \frac{\upd a^\dagger}{\upd t} &= \frac{1}{\I\hbar}[a^\dagger,H] = \I\left(\omega_0 a^\dagger - \omegaab b^\dagger\right) \\
 \label{eq:coupled-f}
 \frac{\upd b^\dagger}{\upd t} &= \frac{1}{\I\hbar}[b^\dagger,H] = \I\left(-\omegaba a^\dagger + \omega_0 b^\dagger\right).
\end{align}
It should be noticed that
\begin{align}
 \frac{\upd a^\dagger}{\upd t} &\neq \left(\frac{\upd a}{\upd t}\right)^\dagger \\
 \frac{\upd b^\dagger}{\upd t} &\neq \left(\frac{\upd b}{\upd t}\right)^\dagger,
\end{align}
that is, lowering and raising operators behave as truly independent dynamical variables in this model. This contrasts with the hermitian formulation, in which the two Dirac operators associated with each quantum oscillator are just hermitian conjugates of each other. This result is not really surprising since $\mathscr{PT}$ symmetry is weaker than hermiticity.

The two sets of coupled operator differential equations~\eqref{eq:coupled-i} and \eqref{eq:coupled-f} may be solved by the canonical transformation
\begin{align}
\label{eq:transformation-i}
 \alpha^- &= \frac{\sqrt{\omegaab} a + \sqrt{\omegaba} b}{\sqrt{2\omegaba}} &
 \beta^- &= \frac{\sqrt{\omegaab} a - \sqrt{\omegaba} b}{\sqrt{2\omegaba}} \\
 \alpha^+ &= \frac{\sqrt{\omegaba} a^\dagger + \sqrt{\omegaba} b^\dagger}{\sqrt{2\omegaab}} &
 \beta^+ &= \frac{\sqrt{\omegaba} a^\dagger - \sqrt{\omegaab} b^\dagger}{\sqrt{2\omegaab}}.
\label{eq:transformation-f}
\end{align}
These new operators are the lowering ($\alpha^-$ and $\beta^-$) and the raising ($\alpha^+$ and $\beta^+$) operators for the two intercavity field modes. It should be remarked, however, that the intercavity field modes are delocalized, consisting of a superposition of localized cavity modes from both cavities.  Nonlocal hermitian interactions may, in some cases, be described more simply by a non-hermitian hamiltonian~\cite{Mostafazadeh:2003b}. From now on, we will refer to the description of the system given by the cavity field mode operators $a$, $a^\dagger$, $b$, and $b^\dagger$ as the \emph{localized representation} while the expression \emph{delocalized representation} will be associated to the description of the system given by the intercavity field mode operators $\alpha^-$, $\alpha^+$, $\beta^-$ and $\beta^+$.

\section{Time evolution in the delocalized representation}

The intercavity field mode operators satisfy the expected equal-time canonical commutation relations and their evolution is ruled by the uncoupled differential equations
\begin{align}
\label{eq:alpha_beta-i}
 \frac{\upd}{\upd t}\alpha^- &= -\I\omega_\alpha \alpha^- &
 \frac{\upd}{\upd t}\beta^- &= -\I\omega_\beta \beta^- \\
 \frac{\upd}{\upd t}\alpha^+ &= \I\omega_\alpha \alpha^+  &
 \frac{\upd}{\upd t}\beta^+ &= \I\omega_\beta \beta^+,
\label{eq:alpha_beta-f}
\end{align}
where
\begin{equation}
 \omega_\alpha = \omega_0 - \raizabba \qquad \omega_\beta = \omega_0 + \raizabba
\end{equation}
are the eigenfrequencies of the intercavity field modes $\alpha$ and $\beta$, respectively. Equations~\eqref{eq:alpha_beta-i}--\eqref{eq:alpha_beta-f} are solved straightforwardly:
\begin{align}
\label{eq:alphabeta-i}
 \alpha^- &= \alpha^-_0 \e^{-\I\omega_\alpha t} & \beta^- &= \beta^-_0 \e^{-\I\omega_\beta t} \\
 \alpha^+ &= \alpha^+_0 \e^{\I\omega_\alpha t} &
 \beta^+ &= \beta^+_0 \e^{\I\omega_\beta t}.
\label{eq:alphabeta-f}
\end{align}

In the delocalized representation the hamiltonian is quite simple:
\begin{equation}
 H = \hbar\left(\omega_\alpha N^\alpha + \omega_\beta N^\beta + \omega_0 \right)
\end{equation}
where the excitation number operators for the intercavity field modes are given by
\begin{align}
 N^\alpha &= \alpha^+ \alpha^- = \alpha^+_0 \alpha^-_0\\
 N^\beta &= \beta^+ \beta^- = \beta^+_0 \beta^-_0.
\end{align}
As expected, the number of excitations in each intercavity field mode does not depend on time.

Since the excitation number operators for the intercavity field modes are compatible observables, $\ket{n_\alpha,n_\beta}$ is an eigenstate of the hamiltonian with energy
\begin{equation}
 E_{n_\alpha, n_\beta} = \hbar\left(\omega_\alpha n_\alpha + \omega_\beta n_\beta + \omega_0 \right).
\end{equation}
In the regime of weak to moderate intercavity coupling ($\raizabba \leq \omega_0$), energy is always positive.  However, in the strong intercavity coupling regime ($\raizabba > \omega_0$), $E_{n_\alpha, n_\beta}$ may be negative, signalling a breakdown of the rotating-wave approximation. Beside the set $\{N^\alpha, N^\beta\}$, another complete set of operators for the system is formed by the total excitation number operator $N = N^\alpha + N^\beta$ and by the excitation imbalance operator $\Delta_{\alpha\beta} = N^\alpha - N^\beta$. In terms of these operators, the hamiltonian may be cast in the form
\begin{equation}
 H = \hbar\omega_0 \left(N + 1\right) - \hbar\raizabba\Delta_{\alpha\beta}.
\end{equation}
We could not find any evidence of a broken $\mathscr{PT}$-symmetry phase in the spectrum of the eigenvalues of $H$. In fact, it is not clear whether or not this system is required to present a broken $\mathscr{PT}$-symmetry phase as it does not seem to fit in any of the known universality classes discussed by Schomerus~\cite{Schomerus:2011}.

Although the delocalized representation based on intercavity field mode operators is very useful in the theoretical description of the coupled system, the delocalized observables required may not be very easy to measure compared to their localized counterparts $N^\mathrm{A} = a^\dagger a$ and $N^\mathrm{B} = b^\dagger b$ which represent the number of photons in each cavity.

In terms of the localized cavity operators, the total excitation number and the excitation imbalance operators are simply
\begin{align}\label{eq:N}
 N &= a^\dagger_0 a_0 + b^\dagger_0 b_0 \\
 \label{eq:delta}
 \Delta_{\alpha\beta} &= \sqrt{\bafracab} a^\dagger_0 b_0 + \sqrt{\abfracba} a_0 b^\dagger_0.
\end{align}
In the next section, we will investigate the time evolution of the operators of the localized representation.

\section{Time evolution in the localized representation}

The cavities lowering and raising operators are determined from equations~\eqref{eq:alphabeta-i}--\eqref{eq:alphabeta-f}
\begin{align}
\label{eq:ab-i}
 a &= \left(\cos(\sqrt{\abtimesba} t) a_0 +\I\sqrt{\bafracab}\sin(\sqrt{\abtimesba}t) b_0 \right)\e^{-\I\omega_0 t}  \\
 b &= \left(\I\sqrt{\abfracba}\sin(\sqrt{\abtimesba} t) a_0 + \cos(\sqrt{\abtimesba} t) b_0 \right)\e^{-\I\omega_0 t} \\
 a^\dagger &= \left(\cos(\sqrt{\abtimesba} t) a^\dagger_0 -\I\sqrt{\abfracba}\sin(\sqrt{\abtimesba}t) b^\dagger_0 \right)\e^{\I\omega_0 t}  \\
 b^\dagger &= \left(-\I\sqrt{\bafracab}\sin(\sqrt{\abtimesba} t) a^\dagger_0 + \cos(\sqrt{\abtimesba} t) b^\dagger_0 \right)\e^{\I\omega_0 t}.
\label{eq:ab-f}
\end{align}
As a consistency check, we should point out that the equal-time canonical commutation relation~\eqref{eq:ab-commutation1} still holds at any time in this non-hermitian model with non-reciprocal interaction.

%In terms of the operators for the localized representation, the excitation number operators for the intercavity field modes are
%\begin{align}
% N^\alpha &= \frac{1}{2}\left( a^\dagger_0 a_0 + b^\dagger_0 b_0 + \sqrt{\bafracab} a^\dagger_0 b_0 + \sqrt{\abfracba} a_0 b^\dagger_0 \right) \\
% N^\beta &= \frac{1}{2}\left( a^\dagger_0 a_0 + b^\dagger_0 b_0 - \sqrt{\bafracab} a^\dagger_0 b_0 - \sqrt{\abfracba} a_0 b^\dagger_0 \right).
%\end{align}

Within the localized representation, the photon number operators $N^\mathrm{A} = a^\dagger a$ and  $N^\mathrm{B} = b^\dagger b$ for cavities $A$ and $B$, respectively, are incompatible with the hamiltonian, and, therefore, varies in time, as expected. It is simple to show that
\begin{multline}\label{eq:Na}
 N^\mathrm{A} = \cos^2(\sqrt{\abtimesba} t)a^\dagger_0 a_0 + \sin^2(\sqrt{\abtimesba} t)b^\dagger_0 b_0 \\
  +\frac{\I}{\sqrt{\abtimesba}}\cos(\sqrt{\abtimesba} t)\sin(\sqrt{\abtimesba} t)\times \\
  (\omegaba a^\dagger_0 b_0 -\omegaab a_0 b^\dagger_0),
\end{multline}
and that
\begin{multline}\label{eq:Nb}
 N^\mathrm{B} = \sin^2(\sqrt{\abtimesba} t)a^\dagger_0 a_0 + \cos^2(\sqrt{\abtimesba} t)b^\dagger_0 b_0 \\
  +\frac{\I}{\sqrt{\abtimesba}}\cos(\sqrt{\abtimesba} t)\sin(\sqrt{\abtimesba} t)\times \\
  (-\omegaba a^\dagger_0 b_0  + \omegaab a_0 b^\dagger_0).
\end{multline}
Besides, the average photon number in the cavities behaves as expected. For cavity $A$, we have
\begin{multline}
 \bracket{n_\mathrm{A}, n_\mathrm{B}}{N^\mathrm{A}}{n_\mathrm{A}, n_\mathrm{B}} = n_\mathrm{A}\cos^2(\raizabba t) \\ + n_\mathrm{B}\sin^2(\raizabba t)
\end{multline}
while, for cavity $B$,
\begin{multline}
 \bracket{n_\mathrm{A}, n_\mathrm{B}}{N^\mathrm{B}}{n_\mathrm{A}, n_\mathrm{B}} = n_\mathrm{A}\sin^2(\raizabba t) \\ + n_\mathrm{B}\cos^2(\raizabba t).
\end{multline}
In fact, since the system is not coupled to a reservoir,
\begin{equation}
 N = N^\mathrm{A} + N^\mathrm{B} = N^\mathrm{A}_0 + N^\mathrm{B}_0 = a^\dagger_0 a_0 + b^\dagger_0 b_0,
\end{equation}
that is, at any time, the number of photons in the coupled cavities is always equal to the initial number of photons in the system, as expected. The same conclusion is achieved from an analysis of equation~\eqref{eq:N} for the total excitation number operator.

\section{Computation of probability amplitudes}

In the localized representation, the action of the lowering operator $a$ for cavity $A$ on a localized number state $\ket{n_\mathrm{A},n_\mathrm{B}}$ is given by
\begin{multline}
 a\ket{n_\mathrm{A} ,n_\mathrm{B} } = \biggl(\sqrt{n_\mathrm{A} }\cos(\sqrt{\abtimesba} t)\ket{n_\mathrm{A} -1,n_\mathrm{B} } \\  -\I\sqrt{\bafracab}\sqrt{n_\mathrm{B} }\sin(\sqrt{\abtimesba} t)\ket{n_\mathrm{A} ,n_\mathrm{B} -1}\biggr)\e^{-\I\omega_0 t}
\end{multline}
whereas the action of the raising operator $a^\dagger$ for cavity $A$ is
\begin{multline}
 a^\dagger\ket{n_\mathrm{A} ,n_\mathrm{B} } = \biggl(\sqrt{n_\mathrm{A} +1}\cos(\sqrt{\abtimesba} t)\ket{n_\mathrm{A} +1,n_\mathrm{B} } \\  +\I\sqrt{\abfracba}\sqrt{n_\mathrm{B} +1}\sin(\sqrt{\abtimesba} t)\ket{n_\mathrm{A} ,n_\mathrm{B} +1}\biggr)\e^{\I\omega_0 t}
\end{multline}
which are both quantum superpositions in the localized representation. Similar equations hold for $b$, and for $b^\dagger$. In fact, it is worth pointing that
\begin{equation}
\begin{split}
 H_\mathrm{I}\ket{n_\mathrm{A} ,n_\mathrm{B} } &= -\hbar\omegaab\sqrt{n_\mathrm{A} (n_\mathrm{B} +1)}\ket{n_\mathrm{A} -1, n_\mathrm{B} +1} \\  & -\hbar\omegaba\sqrt{(n_\mathrm{A} +1)n_\mathrm{B} }\ket{n_\mathrm{A} +1, n_\mathrm{B} -1}.
\end{split}
\end{equation}
As a consistency check, we note that the vacuum state $\ket{\mathrm{vac}}=\ket{0,0}$ is preserved by the interaction hamiltonian. Another interesting situation arise when the initial state is a pure cavity number state characterized by the same number of photons in both cavity. In this case, the interaction hamiltonian maps the state $\ket{n,n}$ into an entangled state of the $\ket{n+1,n-1}$ and $\ket{n-1,n+1}$ states.

Since the free hamiltonian $H_0$ commutes with the interaction hamiltonian $H_\mathrm{I}$, we may use the Baker--Campbell--Hausdorff formula to obtain
\begin{equation}
 \mathscr{U} = \exp(-\I H t/\hbar) = \exp(-\I H_0 t/\hbar)\exp(-\I H_\mathrm{I} t/\hbar).
\end{equation}
Unfortunately,
\begin{equation}
 [a b^\dagger , a^\dagger b] = N^\mathrm{B} - N^\mathrm{A}
\end{equation}
is not a scalar. Since
\begin{align}
 [a b^\dagger, [a b^\dagger , a^\dagger b]] &= -2 a b^\dagger\\
 [a^\dagger b, [a b^\dagger , a^\dagger b]] &= 2 a^\dagger b,
\end{align}
we would have to consider higher order terms in the Baker--Campbell--Hausdorff formula, which is cumbersome. We turn for the small-time expansion of $\exp(-\I H_\mathrm{I} t/\hbar)\approx 1 - \I H_\mathrm{I} t/\hbar$ around the initial time $t=0$, instead. This approximation suffices in order to draw a physically meaningful picture of the interaction process.

%Apart from a phase, the matrix element of $H_\mathrm{I}$ between two pure number state is
%\begin{equation}
%\begin{split}
% \bracket{n'_\mathrm{A}, n'_\mathrm{B}}{H_\mathrm{I}}{n_\mathrm{A} , n_\mathrm{B} } &= \hbar\omegaab\sqrt{n_\mathrm{A} (n_\mathrm{B} + 1)}\delta_{n'_\mathrm{A}, n_\mathrm{A} - 1}\delta_{n'_\mathrm{B}, n_\mathrm{B} + 1} \\
% & + \hbar\omegaba\sqrt{(n_\mathrm{A} + 1) n_\mathrm{B} }\delta_{n'_\mathrm{A}, n_\mathrm{A} + 1}\delta_{n'_\mathrm{B}, n_\mathrm{B} - 1},
%\end{split}
%\end{equation}
Apart from a phase, the matrix elements of the evolution operator $\mathscr{U}$ are
\begin{equation}\label{eq:bracket_time_interaction}
\begin{split}
 \bracket{n'_\mathrm{A}, n'_\mathrm{B}}{\mathscr{U}}{n_\mathrm{A}, n_\mathrm{B} } &\approx \delta_{n'_\mathrm{A}, n_\mathrm{A} }\delta_{n'_\mathrm{B}, n_\mathrm{B} } \\ & - \I t \biggl(\omegaab\sqrt{n_\mathrm{A} (n_\mathrm{B} + 1)}\delta_{n'_\mathrm{A}, n_\mathrm{A} -1}\delta_{n'_\mathrm{B}, n_\mathrm{B} + 1} \\
 & \quad + \omegaba\sqrt{(n_\mathrm{A} + 1) n_\mathrm{B} }\delta_{n'_\mathrm{A}, n_\mathrm{A} + 1}\delta_{n'_\mathrm{B}, n_\mathrm{B} - 1}\biggr).
\end{split}
\end{equation}
Close inspection of Eq.~\eqref{eq:bracket_time_interaction} reveals that the probability amplitude for a photon to be transmitted from cavity $A$ to cavity $B$ is proportional to $\omegaab\sqrt{n_\mathrm{A} (n_\mathrm{B} + 1)}$ while the probability amplitude for a photon to be transmitted back from cavity $B$ to cavity $A$ is proportional to $\omegaba\sqrt{(n_\mathrm{A} + 1) n_\mathrm{B}}$. Since these amplitudes are, in general, not the same, we may say that the quantum system described in this paper is asymmetrical regarding the exchange of photons. This asymmetry is a consequence of  $\mathscr{PT}$-symmetry in a quantum system.  It must be remarked that this kind of asymmetry is impossible in a closed system described by a hermitian hamiltonian.

\section{Discussion}
\label{sec:discussion}

We proposed a non-hermitian $\mathscr{PT}$-symmetric model for describing the light field of two cavities interacting through a $2D$-chiral mirror. Analysing the time evolution of the system, we showed that the probability amplitude for the exchange of a photon from one cavity to another through the $2D$-chiral mirror differs from the probability amplitude for the reverse process. This result suggests a new possible application of non-hermitian quantum theories, specially $\mathscr{PT}$-symmetric quantum mechanics, as an effective theory dealing with some quantum aspects of light propagation through optical metamaterials. Besides, we must point out the fact that, in this model, the Dirac operators behave as truly independent degrees of freedom, and not just the hermitian conjugates of each other.

\acknowledgments
We acknowledge an anonymous referee for comments that improved the paper and Fr. Saboia de Medeiros Ignatian Educational Foundation for continuous support.

\end{document}